\documentclass[prd,aps,tightline,twocolumn,nofootinbib]{revtex4}
\usepackage{graphicx}
\usepackage{amssymb}
\usepackage{amsmath}
\usepackage{color}

\begin{document}
 
\newcommand{\be}{\begin{equation}}
\newcommand{\ee}{\end{equation}}
\newcommand{\aprime}{\mathbf{a}^{\prime}}
\newcommand{\bprime}{\mathbf{b}^{\prime}}
\newcommand{\kh}{\hat{k}}
\newcommand{\Ip}{\vec{I}_+}
\newcommand{\Imi}{\vec{I}_-}
\newcommand{\bc}{\begin{cases}}
\newcommand{\ec}{\end{cases}}
\newcommand{\cD}{\mathcal{D}}
\newcommand{\xv}{\mathbf{x}}
\newcommand{\qv}{\mathbf{q}}
\newcommand{\pv}{\mathbf{p}}
\newcommand{\trec}{t_{\mathrm{rec}}}
\newcommand{\trei}{t_{\mathrm{rei}}}

\newcommand{\red}{\color{red}}
\newcommand{\cyan}{\color{cyan}}
\newcommand{\blue}{\color{blue}}
\newcommand{\magenta}{\color{magenta}}
\newcommand{\yellow}{\color{yellow}}
\newcommand{\green}{\color{green}}
\newcommand{\rem}[1]{{\bf\blue #1}}

\def\EQ#1{Eq.~(\ref{#1})}
\def\EQS#1{Eqs.~(\ref{#1})}
\def\REF#1{(\ref{#1})}
\def\GEV#1{10^{#1}{\rm\,GeV}}
\def\MEV#1{10^{#1}{\rm\,MeV}}
\def\KEV#1{10^{#1}{\rm\,keV}}
\def\lrf#1#2{ \left(\frac{#1}{#2}\right)}
\def\lrfp#1#2#3{ \left(\frac{#1}{#2} \right)^{#3}}
\def\oten#1{ {\mathcal O}(10^{#1})}
\def\oone{ {\mathcal O}(1)}



\title{
Cosmological Moduli Problem in Low Cutoff Theory 
}


\author{Kazunori Nakayama$^{(a,b)}$,
Fuminobu Takahashi$^{(c,b)}$ and
Tsutomu T. Yanagida$^{(a,b)}$
}

\affiliation{%
$^a$Department of Physics,
     University of Tokyo, Tokyo 113-0033, Japan\\
$^b$Institute for the Physics and Mathematics of the Universe, 
     University of Tokyo, Kashiwa, Chiba 277-8568, Japan\\
$^c$Department of Physics, Tohoku University, Sendai 980-8578, Japan
}

\date{\today}

\vskip 1.0cm

\begin{abstract}
We show that the cosmological moduli problem is solved,
without relying on huge late-time entropy production,
 if the universal cutoff scale of the theory
is a few orders of magnitude smaller than the Planck scale.
We obtain a general estimate of the modulus abundance in terms of the inflationary scale 
and the reheating temperature, and find in particular that the reheating temperature can be
high enough for the  non-thermal leptogenesis to work. 
\end{abstract}

 \maketitle

{\it Introduction : }
Supergravity and string theory are a plausible candidate for an underlying high-energy theory
beyond the standard model. However, these theories suffer from a serious cosmological problem:
the cosmological Polonyi/moduli problem~\cite{Coughlan:1983ci,Banks:1993en}.
The Polonyi field is a singlet supersymmetry (SUSY) breaking field  in the gravity mediation. Its mass is of order 
the gravitino mass, $m_{3/2}$,
 and it has Planck-suppressed interactions with the standard model (SM) particles.
Modulus fields generally appear in the compactification of extra dimensions in string theory
and they obtain a mass of order the gravitino mass or heavier, depending on the stabilization mechanism.
Their interactions are also  considered to be suppressed by the Planck scale.
For simplicity we call those fields as moduli (denoted by $\chi$ collectively), unless otherwise
stated. 

The moduli are copiously produced in the early Universe as coherent oscillations. Because of the long lifetime, 
they dominate the energy density of the Universe, thus altering the standard cosmology in contradiction
with observations. 
The cosmological moduli problem, as will be reviewed shortly, is so severe that some of the solutions
require significant modification of the cosmological  scenarios~\cite{Yamamoto:1985rd,Moroi:1999zb,Endo:2006zj}
such as huge late-time entropy production.
Thus some elaborate mechanism is often required to generate the baryon asymmetry 
of the Universe~\cite{Stewart:1996ai,Dimopoulos:1987rk,Kawasaki:2007yy}.

An elegant solution was proposed long ago by Linde~\cite{Linde:1996cx}.
It was shown that,  if the modulus has a large Hubble mass squared of  $c^2 H^2$ 
with $c\gtrsim \mathcal O(10)$, 
the modulus adiabatically follows the temporal potential minimum 
without inducing coherent oscillations. Most importantly, there is no need for large entropy production,
which makes the leptogenesis scenario viable. 
This adiabatic mechanism was studied in detail recently in Ref.~\cite{Nakayama:2011wqa},
and  it turned out that a small but non-negligible amount of the modulus oscillations
is generically induced at the end of inflation, where the adiabaticity of the modulus dynamics is necessarily violated.
Although the adiabatic solution in its original form does not work,  it may still remain a viable solution
to the moduli problem.

The purpose of this letter is to show that the adiabatic suppression mechanism indeed works in theory with a low cutoff scale,
taking account of the additional modulus production at the end of inflation. We will see that
the moduli problem is actually solved for  relatively low reheating temperature  and the high inflationary scale, for
a wide range of the modulus mass of our interest. 
We believe that this is the simplest solution to the moduli problem.
\\

\begin{figure}[tbp]
\begin{center}
\includegraphics[width=1.0\linewidth]{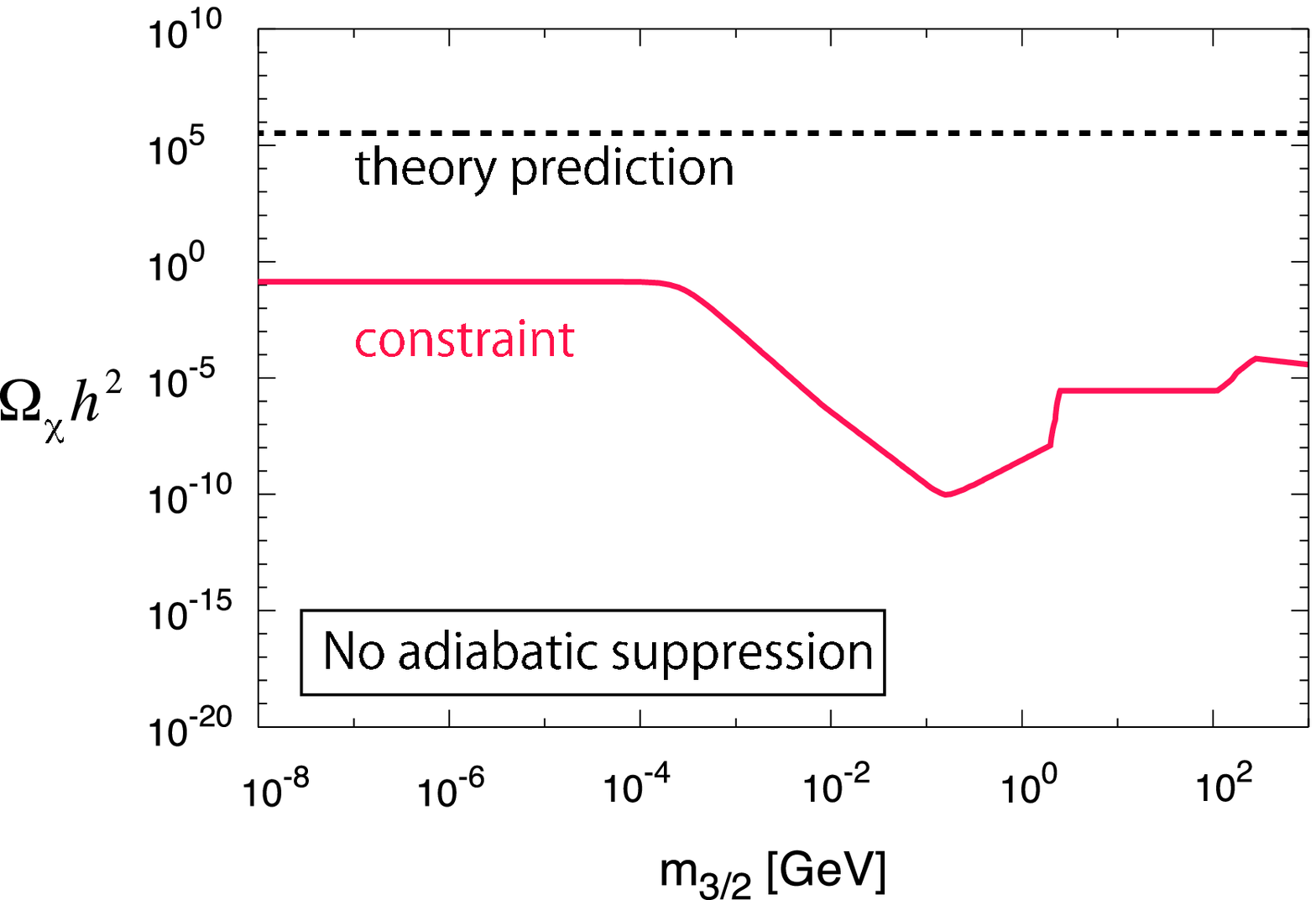}
\includegraphics[width=1.0\linewidth]{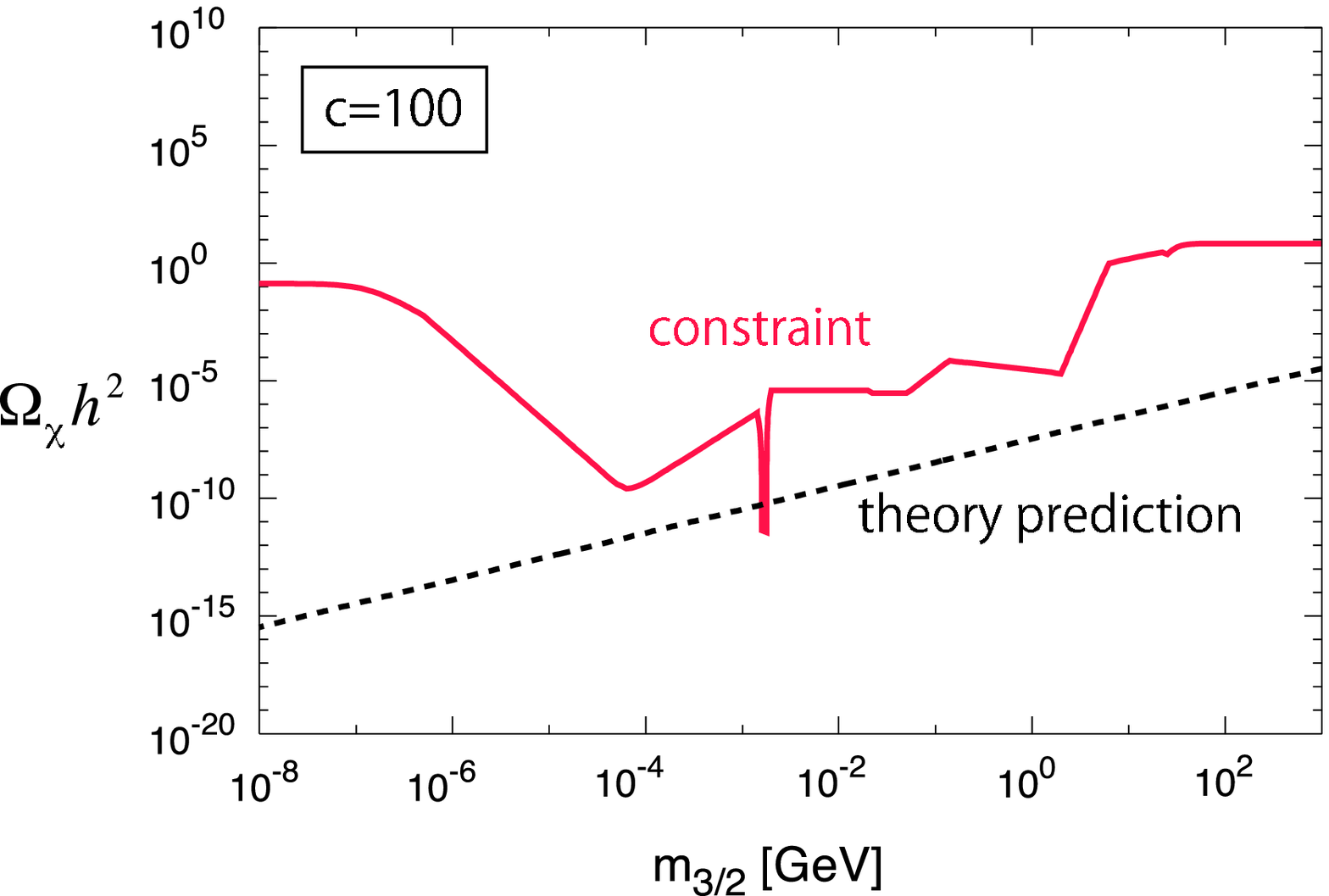}
\caption{
	Upper bound on the modulus abundance as a function of the gravitino mass
	is shown by the red solid line. The top and bottom panels correspond to the cases without and 
	with the adiabatic suppression, respectively. 
	The theoretical predictions (\ref{rhos1}) and (\ref{rhos2}) are also shown by the black dashed line.
          We set $T_{\rm R}=10$\,MeV,  $H_{\rm inf}=10^{13}$\,GeV and $c=100$. Also we set
          $m_\chi = m_{3/2}$ and $\chi_0 = M_P$ in the top panel, and  $m_\chi = c m_{3/2}$ and $\chi_0 = M_P/c$
          in the bottom panel.
}
\label{fig:1}
\end{center}
\end{figure}

{\it Cosmological Moduli Problem : }
Let us briefly review the cosmological moduli problem.
The modulus in general has  a Hubble-induced mass squared during and after inflation,  given by
$m_\chi^{(\rm eff) 2} \sim H^2$. The potential minimum during inflation is generally deviated 
from the true minimum   by $\mathcal O(M_P)$, where $M_P \simeq 2.4 \times \GEV{18}$ is the
reduced Planck mass. When the Hubble parameter becomes 
comparable to the  modulus mass $m_\chi$,
 it begins to oscillate about the true minimum with  an initial amplitude $\chi_0 \sim M_P$.
The ratio of the modulus energy density to entropy density is estimated as
\begin{equation}
	\frac{\rho_\chi}{s}=\frac{1}{8}T_{\rm R}\left( \frac{\chi_0}{M_P} \right)^2
	\simeq 1.3\times 10^5\,{\rm GeV}\left( \frac{T_{\rm R}}{10^6\,{\rm GeV}} \right)
	\left( \frac{\chi_0}{M_P} \right)^2,  \label{rhos1}
\end{equation}
which is related to the density parameter as
$\Omega_\chi h^2 \simeq 2.75\times 10^8 {\rm GeV^{-1}}(\rho_\chi/s)$.
Here we have assumed that the reheating completes after the modulus begins to oscillate, i.e., 
$T_{\rm R} \lesssim \sqrt{m_\chi M_P}$, since otherwise too many moduli or gravitinos
are produced.
Such huge modulus abundance (\ref{rhos1}) is nothing but a cosmological disaster for a wide
range of the modulus mass.
The top panel of Fig.~\ref{fig:1} shows observational constraints on the modulus abundance 
as a function of the gravitino mass, assuming $m_\chi = m_{3/2}$. 
For comparison, the theoretical prediction (\ref{rhos1}) for $T_{\rm R}=10$\,MeV and $\chi_0=M_P$ is shown
as a dashed line.
Observational constraints are based on those described in Ref.~\cite{Asaka:1999xd,Kawasaki:2007mk}
with updated constraints from big-bang nucleosynthesis (BBN)~\cite{Kawasaki:2004qu}.
It is clearly seen that the expected modulus abundance significantly exceeds the observational 
upper bound for the broad range of the modulus mass.
\\

{\it Moduli Problem in Low Cutoff Theory : }
Now let us see how things drastically change in the theory with low cutoff scale.
In fact,  as we shall see below, the adiabatic suppression mechanism is naturally realized
 in this framework~\cite{Takahashi:2010uw}.
Let us take a universal cutoff scale to be $M = M_{P}/c$ with $c\gtrsim \mathcal O(10)$.
Although it is not clear whether all the string-theoretic moduli can have such properties,
at least some of the moduli in the string theory with a relatively large volume of the extra dimensions
compared with the string scale may belong to this class~\cite{hep-th/0605206}.

The Hubble-induced mass for the modulus is enhanced as $m_\chi^{(\rm eff) 2} \sim c^2 H^2$
 by  the following K\"ahler potential,
\begin{equation}
	K \supset \frac{1}{M^2}|\chi|^2 |\Phi|^2,  \label{KNR}
\end{equation}
where $\Phi$ denotes the inflaton field whose $F$-term dominates the Universe during 
and after inflation.
In the case of multi-field inflation model, $\Phi$ can be the waterfall field.
Then the modulus adiabatically follows the temporal potential minimum
without inducing sizable coherent oscillations.  As a result,
the modulus abundance is exponentially suppressed~\cite{Linde:1996cx}.
This is the original version of the adiabatic suppression mechanism.
Notice that since we are considering the theory with lower cutoff scale, $M \sim M_P/c \ll M_P$,
the perturbativity holds up to the scale $M$.
Thus the modulus amplitude can at most have a value of $\sim M$.
Here it should be noted that there also could be the non-renormalizable K\"ahler potential like
$K\sim |\Phi|^4/M^2$ and we need to tune the coefficient of such a term so that it would not spoil the inflaton dynamics.

Recently, we have studied the mechanism in detail,
and found that the a small but non-negligible amount of the modulus oscillations
is generically induced at the end of inflation, where the adiabaticity for the modulus 
is necessarily violated because of the inflaton dynamics~\cite{Nakayama:2011wqa}.
In particular, the abundance is no longer exponentially suppressed.
To be concrete, we consider a multi-field inflation model, in which the inflaton superpotential has a
form of $W = X f(\phi)$. The $X$ takes the role of the inflaton in the hybrid inflation 
model~\cite{Copeland:1994vg,Lazarides:1995vr}, while $\phi$ does in the new inflation model~\cite{Asaka:1999jb}.
To capture the essential features of the supergravity potential with the enhanced coupling \REF{KNR},
we consider the following simplified modulus potential:
\begin{equation}
	V = \frac{1}{2}m_\chi^2 \chi^2 + \frac{1}{2}c_X^2H_1^2 (\chi-\chi_X)^2+\frac{1}{2}c_\phi^2H_2^2 (\chi-\chi_\phi)^2,
	\label{modpot_multi}
\end{equation}
where $c_X, c_\phi \gg 1$, $H_1^2 \simeq (|\dot X|^2+|F_X|^2)/3M_P^2$
and $H_2^2 \simeq (|\dot \phi|^2+|F_\phi|^2)/3M_P^2$. 
The modulus sits at $\chi\sim \chi_X$ during inflation, but it is no longer the exact minimum after inflation ends.
Since the potential minimum changes in the timescale of the inflaton mass, which is much larger than the
modulus mass in most inflation models, the modulus oscillation is induced with the amplitude of 
$\chi_0 \sim (\chi_X-\chi_\phi)c_\phi^2/c_X^2$ just after inflation.
The resultant modulus abundance is given by the following simple formula for multi-field inflation models,
\begin{equation}
	\frac{\rho_\chi}{s} \simeq \frac{1}{8}T_{\rm R}\left( \frac{\chi_0}{M_P/c} \right)^2
	\left( \frac{m_\chi}{cH_{\rm inf}} \right),    \label{rhos2}
\end{equation}
where $H_{\rm inf}$ denotes the Hubble scale during inflation and we have taken $c_\phi=c_X=c$ for simplicity.
This estimate does not depend on the position of the low-energy true minimum of the potential.
Notice that there are two suppression factors compared with (\ref{rhos1}).
One is the smaller initial amplitude $\chi_0\sim M_P/c$ as expected in the low cutoff theory.
The other is a factor of $(m_\chi/H_{\rm inf})$, which gives preference to the small modulus
mass and the high inflation scale.  
On the other hand, for single-field inflation models~\cite{arXiv:hep-ph/9608359},
the modulus potential is given by
\begin{equation}
	V = \frac{1}{2}m_\chi^2 \chi^2 + \frac{V_\phi}{M_P^2}(\chi-\chi')^2+\frac{1}{2}c^2H^2 (\chi-\chi_\phi)^2,
\end{equation}
where $V_\phi$ denotes the inflaton potential energy and the second term comes from the supergravity effect.
(In Eq.~(\ref{modpot_multi}), we have neglected this term because it is subdominant.)
By noting that $V_\phi \simeq 3H^2M_P^2$ during inflation and $\left\langle V_\phi \right\rangle \simeq 
3H^2M_P^2/2$ after inflation, where the bracket represents the time average over the inflaton oscillations,
we find that the modulus potential minimum changes at the end of inflation.
Then the modulus amplitude is given by $3\chi_0/2c^2$ where $\chi_0 = \chi'-\chi_\phi$
and the resulting modulus abundance is
\begin{equation}
	\frac{\rho_\chi}{s} \simeq \frac{9}{32}T_{\rm R}\left( \frac{\chi_0}{M_P/c} \right)^2
	\left( \frac{m_\chi}{c^5 H_{\rm inf}} \right).    \label{rhos2_single}
\end{equation}
This estimate is suppressed by $c^{-4}$ compared to \REF{rhos2}.
To be conservative,  however, we use the estimate of (\ref{rhos2}) in the following.

The couplings of the modulus is also enhanced in theoery with low cutoff scale.
First, the modulus  couples to the SM gauge bosons and gauginos through
\begin{equation}
	\mathcal L = \frac{k}{M}\int d^2 \theta \chi \mathcal W_\alpha \mathcal W^\alpha + {\rm h.c.},
	\label{Zgg}
\end{equation}
where $\mathcal W_\alpha$ denotes the gauge superfield and $k$ is a constant
of order unity.
This induces the modulus decay into the gauge bosons with the rate given by~\cite{Endo:2006ix}
\begin{equation}
	\Gamma(\chi \to gg)\simeq k^2\frac{3m_\chi^3}{2\pi M^2},   \label{Ggg}
\end{equation}
if the modulus decay into all the SM gauge bosons are kinematically allowed.
Notice that the decay rate is enhanced by a factor of $c^5$ compared with the standard case.
The modulus can also decay into gauginos if such modes are kinematically allowed.
Gauginos obtain masses of order $cm_{3/2}$ if there is a singlet SUSY breaking
field. Similarly, sfermions obtain also masses of $cm_{3/2}$. Thus if
$c\simeq O(10)$, the gravitino is the lightest SUSY particle (LSP).

Another important decay mode  is that into gravitinos.
The modulus gets mixed with the SUSY breaking field in general and  the modulus decay rate
into gravitinos  depends on the  SUSY breaking sector and the moduli stabilization.
Assuming that the SUSY breaking field is much heavier than the modulus field and its expectation 
values is negligibly small,\footnote{
	The decay rate into the gravitino is enhanced for the large 
	vacuum expectation value of SUSY breaking field.
}
as in the dynamical SUSY breaking models, we find~\cite{Endo:2006zj}
\begin{equation}
	\Gamma(\chi \to 2\psi_{3/2}) \sim 10^{-3}\frac{m_\chi^3}{M^2}.  \label{grav}
\end{equation}
if the modulus is stabilized by the SUSY mass term.
On the other hand, if the modulus is stabilized by the non-SUSY mass term, the gravitino pair production 
 rate is suppressed. In this case, however, the modulus decays into the gravitino plus modulino, the fermionic 
superpatrner of the modulus,  and its decay rate is comparable to (\ref{grav}). Also, 
if $\chi$ is the Polonyi field, its decay rate into a pair of the gravitinos is given by Eq.~(\ref{grav}).
Thus we adopt (\ref{grav}) as the gravitino production rate.
The gravitino (modulino) can decay into modulino (gravitino) + $2\gamma$ by exchanging the modulus
if the former is heavier than the latter.
The lifetime of this process, however, is sufficiently long so that the emitted gamma-rays do not have
observational consequences.

Taking  those decay modes into account, 
we have estimated observational constraints on the modulus abundance in this framework.
The bottom panel of Fig.~\ref{fig:1} shows the observational bounds on the modulus abundance,
in comparison with the theoretical prediction for $T_{\rm R}=10$\,MeV for $c=100$, 
$\chi_0 = M_P/c$, $m_\chi = c m_{3/2}$, and $H_{\rm inf}=10^{13}$\,GeV.
It is seen that the modulus abundance is far below the observational  bound
for almost all the masses.
Therefore, the moduli problem is solved simply by  making the cutoff scale a few orders of magnitude smaller.
No additional entropy production is needed.

\begin{figure}[tbp]
\begin{center}
\includegraphics[width=0.9\linewidth]{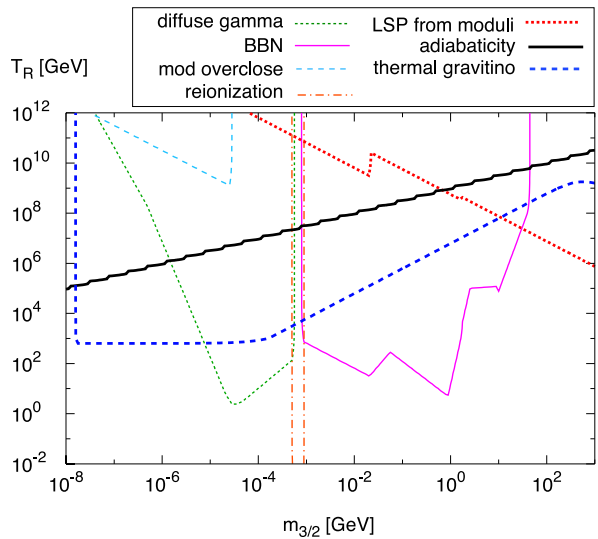}
\caption{
	Upper bound on the reheating temperature as a function of the modulus mass
	for $c=100$, $H_{\rm inf}=10^{13}$\,GeV and $k=1$. 
	Constraints include BBN, diffuse gamma-ray background,
	reionization, thermal gravitino overproduction,
	LSP (gravitino) overproduction from the modulus decay and adiabaticity. 
}
\label{fig:2}
\end{center}
\end{figure}

In Fig.~\ref{fig:2} we show the various upper bounds on the reheating temperature as a function of the gravitino mass.
Note that the gravitino is the LSP. Here the BBN constraint on the decay of the next-lightest-SUSY particle (NLSP)
is not taken into account,  assuming that the NLSP decays quickly via R-parity violating operators,
if the lifetime of the NLSP exceeds $1$ sec. Notice that there is a general upper bound on the reheating temperature 
in order for the adiabatic suppression to work, 
which roughly reads $T_{\rm R}\lesssim 0.05\sqrt{m_\chi M_P}$~\cite{Takahashi:2011as, Nakayama:2011wqa}.
This is shown by the line labeled as ``adiabaticity''.
The bound from the gravitino thermal production is shown as thick dashed (blue) line 
~\cite{Moroi:1993mb,Bolz:2000fu,Pradler:2006qh,Rychkov:2007uq}.
It is seen that the moduli problem is solved for $T_{\rm R}\lesssim 1$\,TeV for almost all the mass range.
There are some parameter spaces where constraints are less severe,  around
$10{\rm\,GeV} \lesssim m_{3/2} \lesssim 1{\rm\,TeV}$.
\\

{\it Implications : }
Our solution to the cosmological moduli problem has some phenomenological implications.
First of all, because of the bound on $T_{\rm R}\lesssim 0.05\sqrt{m_\chi M_P}$,
one of the windows where leptogenesis~\cite{Fukugita:1986hr} works, $m_{3/2} \lesssim \mathcal O(10)$eV, is closed. 
On the other hand, there is a window around $m_{3/2}\sim 100$\,GeV  where all the 
constraints are less severe and the reheating temperature 
as high as $T_{\rm R}\sim 10^7$\,GeV is allowed. Then, 
non-thermal leptogenesis~\cite{Asaka:1999yd,Hamaguchi:2001gw}  is possible.
This is very appealing, since the moduli problem is solved without disturbing a successful leptogenesis scenario.
In this case sfermions may be as heavy as $10$\,TeV, and as a consequence,
the lightest Higgs boson may be as heavy as 120-125\,GeV.
For the other values of the modulus mass, 
 the reheating temperature is constrained as $T_{\rm R} \lesssim$ TeV,
and other baryogenesis scenarios are needed. 
Notice that the bound can easily be relaxed by a few orders of magnitude
by reducing $\chi_0$ slightly or by choosing $\mathcal O(1)$ 
constants in (\ref{Ggg}) and (\ref{grav}) appropriately.

Here are a few remarks.
First, as is clear from the expression (\ref{rhos2}), 
the modulus abundance is inversely proportional to the inflation scale $H_{\rm inf}$.
Thus,  high scale inflation models are favored in this respect.\footnote{
	Note that $H_{\rm inf}$ in Eq.~(\ref{rhos2}) corresponds to the Hubble scale at the end of inflation,
	and not the Hubble scale when observational scales exit the horizon.
}
In supergravity, chaotic inflation models with
a shift symmetry were proposed~\cite{Kawasaki:2000yn,Takahashi:2010ky}.
There are also models in the Jordan frame supergravity~\cite{Einhorn:2009bh,Ferrara:2010yw,Lee:2010hj,Nakayama:2010ga}.
More general arguments are found in recent publications~\cite{Kallosh:2010ug}. Also, 
the chaotic inflation models are  favored since the $Z_2$-symmetry on the inflaton field
can forbid the dangerous non-thermal gravitino overproduction~\cite{Kawasaki:2006gs}.
Second, as  mentioned below Eq.~(\ref{rhos2}), 
the modulus abundance is further  suppressed in single-field inflation models.
Although the known single-field inflation model~\cite{arXiv:hep-ph/9608359}
predicts low inflation scale, $H_{\rm inf}\sim 10^7$\,GeV, the additional suppression factor $c^{-4}$ 
can easily compensate, making the single new inflation model viable in this context.
\\

{\it Conclusions : }
In this letter we have shown that the notorious cosmological moduli problem is
solved if the cutoff scale of the theory is lower than the Planck scale. 
For the cutoff scale of $M\sim 10^{-2}M_P$, the modulus abundance is suppressed enough to be consistent with observations
for almost all the gravitino mass $10$\,eV $\lesssim m_{3/2} \lesssim 1$\,TeV, or equivalently, 
the modulus mass $1{\rm\,keV} \lesssim m_\chi \lesssim 100$\,TeV.
Although the reheating temperature is  constrained from above (see Fig.~\ref{fig:2}) ,
no significant modification of the standard cosmology is needed. In particular, 
for the gravitino mass of $\mathcal O(100)$\,GeV, the reheating temperature can be as high as
$T_{\rm R}\sim 10^7$\,GeV and hence non-thermal leptogenesis is possible.
We also note that the modulus abundance is suppressed for high-energy scale inflation.
Thus high-energy scale inflation models are favored from the viewpoint of cosmological moduli 
problem.
\\

{\it Acknowledgments : }
This work was supported by the Grant-in-Aid for Scientific Research on
Innovative Areas (No. 21111006) [KN and FT], Scientific Research (A)
(No. 22244030 [KN and FT], 21244033 [FT], 22244021 [TTY]), and JSPS Grant-in-Aid for
Young Scientists (B) (No. 21740160) [FT].  This work was also
supported by World Premier International Center Initiative (WPI
Program), MEXT, Japan.


   
 
 \end{document}